\documentclass[twocolumn]{aastex61}

\usepackage{graphicx}	
\usepackage{amsmath}	
\usepackage{amssymb}	
\usepackage{enumitem}
\usepackage{soul}                          
\usepackage{ulem} 
\usepackage{xifthen}
\usepackage{xcolor}
\usepackage{hyperref}	
\hypersetup{colorlinks=true,linkcolor=blue,citecolor=blue,filecolor=blue,urlcolor=blue}

\newcommand{\Msun}{\,{\rm M_\odot}}

\newcommand{\Msunperyr}{\,{\rm M_\odot} \ \rm{yr}^{-1}}
\newcommand{\fedd}{\,{f_{\rm Edd}}}
\newcommand{\Mblack}{M_\bullet}

\newcommand\mybar{\kern1pt\rule[-\dp\strutbox]{.8pt}{\baselineskip}\kern1pt}

\newcommand{\change}[2][]{%
\ifthenelse{\isempty{#2}}{{\color{red}{#1}}}%
{{\color{orange}\sout{#1}}{\color{red}{#2}}}%
}
\setlist[itemize]{noitemsep, topsep=0pt, leftmargin=*}

\shorttitle{Active Fraction of MBHs in Dwarf Galaxies}
\shortauthors{Pacucci, Mezcua \& Regan}

\begin{document}

\title{The Active Fraction of Massive Black Holes in Dwarf Galaxies}
\correspondingauthor{Fabio Pacucci}
\email{fabio.pacucci@cfa.harvard.edu}

\author[0000-0001-9879-7780]{Fabio Pacucci}
\affil{Center for Astrophysics $\vert$ Harvard \& Smithsonian,
Cambridge, MA 02138, USA}
\affil{Black Hole Initiative, Harvard University,
Cambridge, MA 02138, USA}

\author[0000-0003-4440-259X]{Mar Mezcua}
\affil{Institute of Space Sciences (ICE, CSIC), Campus UAB, Carrer de Magrans, 08193 Barcelona, Spain}
\affil{Institut d'Estudis Espacials de Catalunya (IEEC), Carrer Gran Capit\`a, 08034 Barcelona, Spain}

\author[0000-0001-9072-6427]{John A. Regan}
\affil{Department of Theoretical Physics, Maynooth University, Maynooth, Ireland}

\begin{abstract}
\noindent The population of massive black holes (MBHs) in dwarf galaxies is elusive, but fundamentally important to understand the coevolution of black holes with their hosts and the formation of the first collapsed objects in the Universe. While some progress was made in determining the X-ray detected fraction of MBHs in dwarfs, with typical values ranging from $0\%$ to $6\%$, their overall active fraction, ${\cal A}$, is still largely unconstrained. Here, we develop a theoretical model to predict the multiwavelength active fraction of MBHs in dwarf galaxies starting from first principles and based on the physical properties of the host, namely, its stellar mass and angular momentum content. We find multiwavelength active fractions for MBHs, accreting at typically low rates, ranging from $5\%$ to $22\%$, and increasing with the stellar mass of the host as ${\cal A} \sim(\log_{10}M_{\star})^{4.5}$. If dwarfs are characterized by low-metallicity environments, the active fraction may reach $\sim 30\%$ for the most massive hosts.
For galaxies with stellar mass in the range $10^7<M_{\star} [\Msun]<10^{10}$, our predictions are in agreement with occupation fractions derived from simulations and semi-analytical models. Additionally, we provide a fitting formula to predict the probability of finding an active MBH in a dwarf galaxy from observationally derived data. This model will be instrumental to guide future observational efforts to find MBHs in dwarfs. The James Webb Space Telescope, in particular, will play a crucial role in detecting MBHs in dwarfs, possibly uncovering active fractions $\sim 3$ times larger than current X-ray surveys.
\end{abstract}


\section{Introduction} \label{sec:intro}
\noindent Black hole (BH) seeds of mass $10^3 \lesssim \Mblack (\Msun) \lesssim 10^{6}$ formed at redshifts $z > 10$ \citep{BL01} have been invoked to explain the presence of quasars of $\Mblack \sim 10^{9-10} \Msun$ when the Universe was less than 1 Gyr old \citep[$z\sim7$; e.g.][]{Banados_2018, Matsuoka_2018, Wang_2021}. 
These seeds, already born significantly more massive than typical BHs of stellar origins, are likely to have contributed to the formation of supermassive black holes (SMBHs, typically with a mass $\Mblack > 10^6 \Msun$) in massive local galaxies \citep[e.g.][]{McConnell_Ma_2013, Mezcua_2018, Pacucci_2018}. However, those seeds that did not reach the SMBH stage should be observable today in local dwarf galaxies (typically with a stellar mass $M_{\star} < 10^{10} \Msun$), which are those that best resemble the first galaxies formed in the early Universe \citep[][]{VanWassenhove_2010, Pacucci_MaxMass_2017, Ricarte_2018}. These BHs are known as massive black holes (MBHs). In this study, we define MBHs in the mass range $10^3 \lesssim \Mblack (\Msun) \lesssim 10^{7}$, similar to the definition in \cite{Bellovary_2019}. Studying the number of MBHs in local dwarf galaxies, i.e. their BH occupation fraction, could thus help us understand the population of high-z seed BHs if these have not significantly grown via accretion and mergers with cosmic time (e.g. \citealt{Volonteri_2008, Pacucci_MaxMass_2017, Mezcua_2019, Pacucci_Loeb_2019}).

The origin of BH seeds remains unknown, with a number of pathways under intense investigation (see \citealt{Woods_2019, Inayoshi_review_2019} for recent reviews). 
The progenitor BH seeds could be the end product of Population III (Pop III) stars \citep{Madau_Rees_2001, Bromm_2004}, which are expected to leave in their wake a population of BHs with masses $< 1,000 \Msun$ \citep{Hirano_2014}, commonly named light seeds. Alternatively, the progenitors may form from the collapse of massive and pristine gas clouds, leading to heavy seeds with masses up to a few times $10^5 \Msun$ \citep{Loeb_Rasio_1994, Bromm_Loeb_2003, Begelman_2006,Regan_2009, Bromm_Yoshida_2011}. These seeds are born in near-pristine atomic cooling haloes that support high accretion rates ($\dot{M} \gtrsim 10^{-3} \Msunperyr$) onto embryonic protostars. 
Finally, BH seed formation could result from gravitational runaway events inside dense clusters either through stellar collisions \citep{PortegiesZwart_2002, PortegiesZwart_2004, Gurkan_2004, Katz_2015, Boekholt_2018}, through BH mergers \citep{Davies_2011, Lupi_2016, Antonini_2019} or through rapid accretion \citep{Alexander_2014, Natarajan_2021}. 

In each case, the BH seed can then rapidly grow via accretion and subsequent mergers \citep{Pacucci_2020} and reach the observed SMBH masses of $\Mblack \sim 10^{9-10} \Msun$ by $z\sim7$ \citep[see also reviews by][]{Volonteri_2010, Mezcua_2017_review, Greene_review_2020}.
The detection of high-$z$ seeds may be possible with the new generation of observational facilities such as the \textit{James Webb Space Telescope} \citep[JWST, e.g.,][]{PFVD_2015, Natarajan_2017, Valiante_2017, Whalen_2020_JWST}, the \textit{Athena} X-ray observatory \citep{Valiante_2017}, the \textit{Lynx} mission concept \citep[e.g.][]{Ricarte_2018, Mezcua_2018_a}, the Square Kilometer Array \citep[e.g.][]{Whalen_2020} or via gravitational waves using the LISA observatory \citep{LISA_2017, Hartwig_2018, Sesana_2019}.

In the last decade tens of accreting MBHs candidates have been found in dwarf galaxies as low-mass ($\Mblack \lesssim 10^{7} \Msun$) active galactic nuclei (AGN). Most searches have focused on the use of optical narrow emission line diagnostic diagrams to identify AGN emission and of broad emission lines to estimate the BH mass (e.g., \citealt{Greene_Ho_2004, Greene_Ho_2007, Reines_2013, Moran_2014, Chilingarian_2018}). The use of optical variability \citep{Baldassare_2018, Baldassare_2020_PTF, Martinez-Palomera_2020} and integral field spectroscopy (IFS, \citealt{Mezcua_2020_rot_sup}) has yielded several additional tens of ``hidden'' low-mass AGN whose emission is diluted by the star formation of the host galaxy. 
Most of these samples are however biased toward very local ($z<0.3$) dwarf galaxies and are limited by obscuration. This has been circumvented by infrared (e.g., \citealt{Satyapal_2008, Marleau_2017, Secrest_Satyapal_2020}), X-ray (e.g., \citealt{Schramm_2013, Lemons_2015, Mezcua_2016, Mezcua_2018_a, Birchall_2020}), and radio searches \citep{Mezcua_2019_radio, Reines_2020} of AGN in dwarf galaxies, which has yielded the detection of AGN in dwarf galaxies out to $z\sim3.4$ \citep{Mezcua_2019_radio}. The use of wide and deep X-ray surveys has additionally provided a robust measurement of the number of AGN in dwarf galaxies out to $z\sim0.7$ \citep{Mezcua_2018_a, Birchall_2020}. In the near future, eROSITA, with its planned eight full scans of the X-ray sky, could also provide important information about accreting sources \citep{erosita_2021}.

These observational techniques have significantly improved our view of the presence of MBHs in dwarfs, but they intrinsically fall short in the determination of the real BH occupation fraction of dwarfs because they rely on the assumption that the BH needs to be accreting in order to emit radiation. Hence, these techniques probe the \textit{active fraction} of dwarfs, i.e. the fraction of dwarf galaxies hosting an actively accreting BH.

High-resolution hydrodynamic simulations have been complementing observational advances and have recently been used to probe the formation of BH seeds in early galaxies (mostly on the heavy mass range because of dynamic range limitations of the simulations). While a complete understanding of their formation still eludes us, sub-grid models can be used to study their evolution under certain assumptions. 
Different studies use slightly different sub-grid models for seeding BHs in early galaxies with typical initial seed masses between $\Mblack = 10^{4-5} \Msun$. This approach leads to a population of early galaxies with seed BHs that can then undergo accretion and merger events and grow into the population of SMBHs observed today in both galactic centers and as off-nuclear sources \citep{Mezcua_2017_review}. However, this treatment neglects the pathway that led to the BH seed. As a result, the seeding (sub-grid) model used can influence the occupation fraction particularly at masses comparable to the halo masses in which the BHs are seeded (i.e. dwarf galaxies).

For example, \cite{Bellovary_2019} use the \textsc{Changa} code to model the formation and evolution of BHs in early galaxies. They employ a seed mass of $\Mblack \sim 25,000 \Msun$ with initial seeding occurring in haloes with halo mass M$_{h} \gtrsim 10^7 \Msun$. Using this sub-grid model and by examining the subsequent hierarchical structure evolution, they find BH occupation fractions of nearly 80\% for galaxies with stellar masses M$_* \sim 10^9 \Msun$ dropping down to the percent level for galaxies with M$_* \lesssim 10^6 \Msun$. Crucially, \cite{Bellovary_2019} also find that the vast majority of MBHs in their dwarf galaxies are extremely faint and may be difficult to detect electromagnetically. \cite{Ricarte_2021}, however, point out that this population of so-called ``wandering MBHs'' may contribute significantly to detection rates of tidal disruption events. Nonetheless, these conclusions would then explain the low active fraction of AGN thus far discovered through observations. \textit{These findings highlight the crucial difference between occupation and active fractions}.

While high-resolution hydrodynamic simulations provide excellent insight into the growth and evolution of structure, including BHs, they come at a substantial computational cost. This cost limits their ability to explore large parameter spaces, which is one of the reasons why sub-grid modeling of BH formation is used. Instead, semi-analytical models (SAMs) can be used to explore a much larger parameter space. 
SAMs largely use dark matter merger trees to backbone their models with the inclusion of baryonic physics, including BH formation, constructed using formulations based on both empirical and theoretical models \citep[][]{Valiante_2017, Barausse_2012, Pacucci_2018, Dayal_2019, Barausse_2020}. As a result, SAMs are unable to directly follow nonlinear processes like galaxy mergers and BH mergers but they are extremely useful for exploring large sections of the parameter space.

\indent \cite{Ricarte_2018} used their SAMs to explore what observational signatures may be able to discriminate between seeding models and accretion modes. In particular, they find that signatures of the BH seeding can be found in the occupation fractions and the luminosity functions of high-redshift SMBHs. In their models, the occupation fraction is essentially unity for BHs in haloes with stellar masses of $M_{\star} \gtrsim 10^{10} \Msun$, dropping to the percent level for haloes with stellar masses of M$_* \sim 10^{8} \Msun$. Perhaps more importantly, however, they find that the X-ray luminosity function of accreting BHs is potentially a powerful probe of the seeding mechanisms, with the luminosity function displaying distinctly different features between light and heavy BHs that becomes more pronounced with increasing redshift. We compare our predictions to the predictions on the occupation fraction of \cite{Ricarte_2018} in \S \ref{subsec:fractions}.

Fundamentally, our work follows a different approach to study the active fraction of MBHs in dwarf galaxies. We employ a theoretical model for BH growth that takes into account the physical properties of the BH and its host environment, i.e. the mass of the dwarf and the angular momentum content in its innermost regions. Our physical model can be used to predict multiwavelength active fractions of MBHs in dwarfs. 

Our model is complementary to, and is compared against, cosmological simulations and SAMs, and provides a theoretical framework that can be used to explore a very large phase space at minimal computational cost. In contrast to both hydrodynamical simulations and SAMs, our model does not follow the evolution of individual haloes but instead calculates MBH growth rates based on the physical conditions of a galaxy. In the following \S \ref{sec:theory} we introduce and describe our theoretical model in detail and in \S \ref{sec:calibration} we describe its calibration against X-ray data. Finally, in \S \ref{sec:results} and \S \ref{sec:disc_concl} we describe the results and sum up our conclusions. All results are presented at $z \sim 0$, unless otherwise stated. \\

\section{Theoretical Model} 
\label{sec:theory}
We first provide a clear definition of three concepts that we frequently refer to in this paper: occupation fraction, active fraction, and X-ray detected fraction.
\begin{itemize}
    \item \textbf{Occupation fraction:} the fraction of dwarf galaxies that contain an MBH. It can be active or dormant, central, or noncentral (see, e.g., \citealt{Bellovary_2019, Greene_review_2020}).
    \item \textbf{Active fraction, ${\cal A}$:} the fraction of dwarf galaxies that contain an accreting MBH. In our implementation, this translates into a duty cycle, $\cal{D}$, (fraction of total time spent accreting) close to $\sim 1$ and Eddington ratio $f_{\rm Edd} = \dot{M}/\dot{M}_{\rm Edd} \sim 0.1$, typical for low-$z$ AGN (see, e.g., \citealt{Greene_Ho_2007, Suh_2015}). In other words, the BH is accreting with a typical accretion rate for low-$z$ sources around the observation time. 
    \item \textbf{X-ray detected fraction:} the fraction of dwarf galaxies that contain an MBH that is detected in the X-ray and whose emission can be clearly distinguished from the high-energy emission of other sources, e.g. X-ray binaries (see, e.g., \citealt{Mezcua_2018, Birchall_2020}).
\end{itemize}
\vspace{2pt}
\textit{Our goal is to predict the active fraction of MBHs in dwarf galaxies, as a function of \underline{both} the stellar mass of the host and of the parameter of rotational support of gas inside the nucleus.} The parameter of rotational support, properly defined in \S \ref{subsec:observables}, is used as a proxy of the angular momentum content of the gas close to the MBH.

A methodological note: in our model, efficient accretion occurs only for central BHs, as the compact object needs to be in the deepest section of the galactic potential well to have a large gas reservoir at its disposal. Recent studies (e.g., \citealt{Bellovary_2019, Bellovary_2021, Ricarte_2021}) have suggested that a large fraction of MBHs in galaxies can be off-centered, with perhaps up to $\sim 50\%$ of MBHs in low-mass galaxies off-center \citep{Bellovary_2019}. Although the observation of these compact objects is challenging \citep{Bellovary_2021} due to the paucity of available gas to accrete, the possibility that the off-nuclear MBHs are accreting cannot be ruled out, and indeed several examples have recently been found in dwarf galaxies (see, e.g., \citealt{Mezcua_2018_off, Reines_2020, Mezcua_2020_rot_sup}). In our fiducial model we do not include off-centered MBHs as they likely have very low accretion rates \citep{Bellovary_2019}.

\subsection{Modeling Efficient BH Growth}
\label{subsec:BH_growth}
We use a model for efficient accretion onto central BHs very similar to the one developed in \cite{Pacucci_2017}, to which the interested reader is referred to for a detailed description. We define an efficient accretion flow for a low-$z$ AGN as one with a duty cycle ${\cal D} \sim 1$ and a typical Eddington ratio $\fedd \sim 0.1$ (see, e.g., \citealt{Greene_Ho_2007, Suh_2015}). This ideal situation can arise when the combined effects of the angular momentum and of the radiation pressure are ineffectual in suppressing the accretion flow, at large and small spatial scales. The reference spatial scale, to discriminate between small and large, is set at the Bondi radius $R_B = 2 G M_{\bullet}/c_s^2$, (\citealt{Bondi_1952}, where $M_{\bullet}$ is the BH mass and $c_s$ is the sound speed of the unperturbed gas around the BH).

The model described below is based on two input quantities: the number density of the gas at the Bondi radius, $n_{B}$, and the angular momentum content at the Bondi radius, parameterized as $\lambda_{\rm B} = \ell_{\rm B}/(GMR_{\rm B})^{1/2}$, which is the ratio of the angular momentum per unit mass of gas, $\ell_{\rm B}$, and its value for a circular orbit (at $R_B$, see \citealt{Begelman_Volonteri_2017}).
Of course, these two quantities are not easily determined with observations. In \S \ref{subsec:observables} we derive some proxies for $n_{B}$ and $\lambda_{\rm B}$ which can be more readily determined from observations.

The condition on the BH mass $\Mblack$ for efficient growth on \textit{small scales}, $r \ll R_B$, is:
\begin{equation}
    \Mblack \gtrsim M_{\bullet,S} = 10^{-1}  \left(\frac{n_{B}}{10^5 \, \mathrm{cm^{-3}}}\right)^2 \, \mathrm{\Msun} \, ,
\label{eq:cond1}
\end{equation}
and depends on $n_{B}$.
On large scales, $r \geq R_B$, the relevant condition depends on both $n_{B}$ and $\lambda_{\rm B}$, as well as the Eddington ratio $\fedd$ which here is explicitly expressed:
\begin{equation}
    \Mblack \gtrsim M_{\bullet,L} = \frac{3.6 \times 10^{13}}{\sqrt{\fedd}} \left( \frac{n_{B}}{\mathrm{10^5 \, cm^{-3}}} \right)^{-1} \lambda_{\rm B}^{2} \,  \mathrm{\Msun} \, .
\label{eq:cond2}
\end{equation}

As Eqs. \ref{eq:cond1} and \ref{eq:cond2} show, the BH mass is required to be larger than a minimum mass which depends on the gas number density and on the nondimensional angular momentum parameter, both computed at the Bondi radius. As the dependence on $\fedd$ is only through its square root, our results are not heavily dependent on the specific value of the typical Eddington ratio we choose. We set its value to $0.1$ as typical for low-$z$ AGN (see, e.g., Fig. 1 of \citealt{Greene_Ho_2007} and the lowest redshift bin in Fig. 5 of \citealt{Suh_2015}), accreting in an environment where cold gas is already scarce \citep{Power_2010}. Also, note that $\fedd$ is the Eddington ratio of the inflow of gas, not necessarily the rate at which the BH is ultimately growing.

\begin{figure}[h]
\includegraphics[angle=0,width=0.49\textwidth]{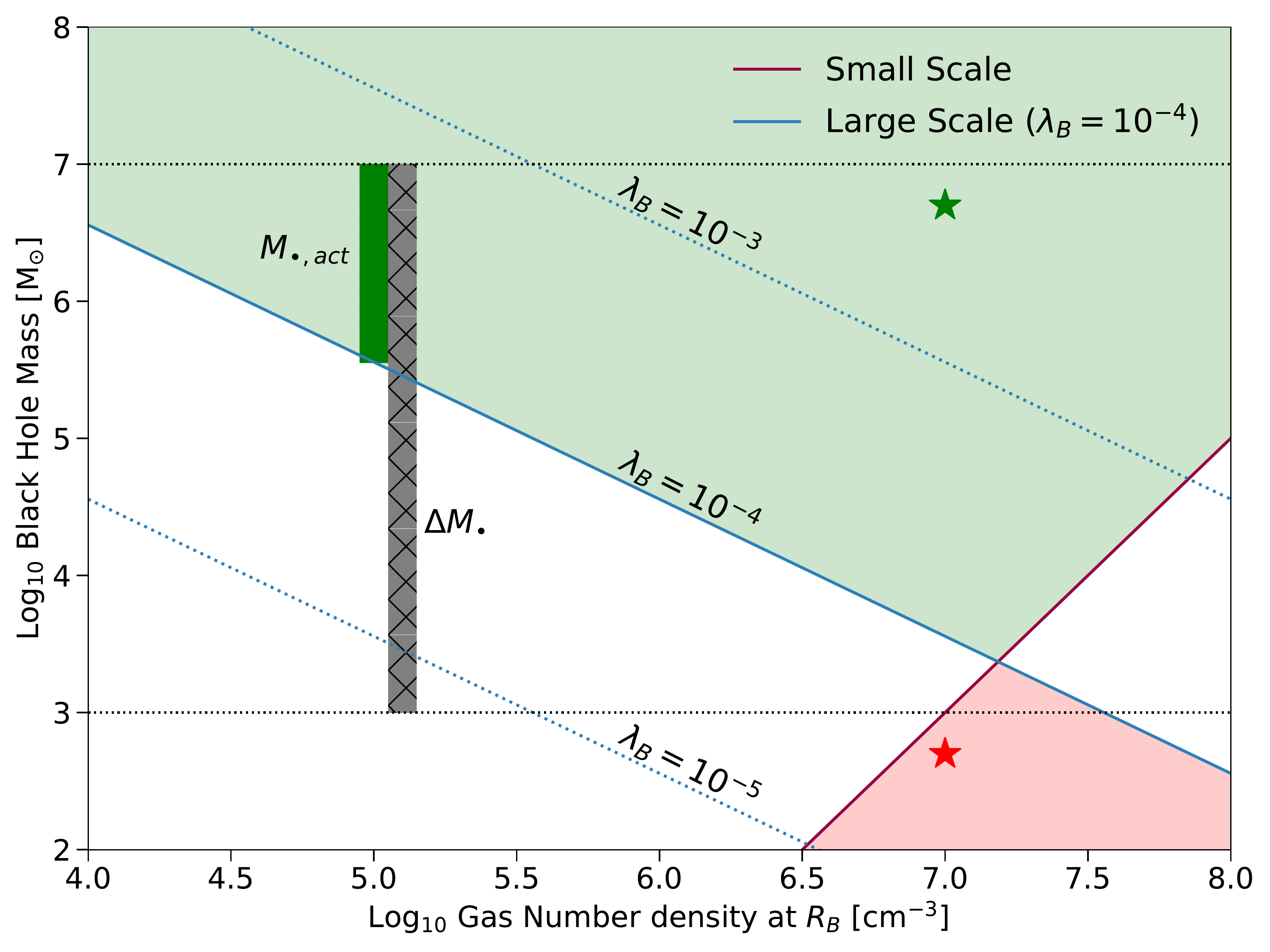}
    \caption{Example application of the model for efficient accretion. The solid lines indicate the conditions expressed in Eqs. \ref{eq:cond1} and \ref{eq:cond2} (for $\lambda_B = 10^{-4}$). The green area is the most efficient region for accretion, while the red is the most inefficient one. For example, a BH located where the green star is would accrete efficiently, and conversely for the red star. The dotted lines indicate the condition in Eq. \ref{eq:cond2} for two additional values of $\lambda_B$, as indicated, for reference. To calculate the active fraction we calculate the ratio of the BH mass range that allows for efficient accretion compared to the total MBH mass range. For instance, the green hatched box indicates the value of $M_{\bullet, act}$ for the density value of $10^5 \, \mathrm{cm^{-3}}$, while the gray hatched box indicates the whole mass range $\Delta \Mblack$ - see \S \ref{subsec:act_frac} for details. Figure adapted from \cite{Pacucci_2017}.}
    \label{fig:galaxy_BH}
\end{figure}

If a BH is to accrete efficiently, this model assumes that its mass has to be larger than both the limits indicated in Eqs. \ref{eq:cond1} and \ref{eq:cond2}. The BH mass $\Mblack$ range that verifies both conditions is indicated by the green region in Figure \ref{fig:galaxy_BH}. We place an upper limit of $\Mblack = 10^7 \Msun$ equal to the typical BH mass found in dwarf galaxies of $10^{10} \Msun$ \citep{Greene_review_2020}.
Figure \ref{fig:galaxy_BH} exemplifies this model in a simple case of two BHs, one in the efficient region and one in the inefficient one. Note that gas densities of $\sim 10^4-10^7 \, \mathrm{cm^{-3}}$ are found in gas clouds at the center of the Milky Way \citep{Mills_2018}.

\subsection{Matching to Observables}
\label{subsec:observables}

The equations used to identify the conditions for efficient accretion (Eqs. \ref{eq:cond1}, \ref{eq:cond2}) are expressed as a function of the number density of the gas and of its angular momentum content, both computed at the Bondi radius. These quantities are not easily derived from observations, hence in this subsection we show how to express them via proxies.

First, we approximate the number density of the gas in terms of the stellar mass of the host. In this way, we are able to estimate the probability that an MBH of mass $\Mblack$ is within the efficient accretion region in Fig. \ref{fig:galaxy_BH}, and hence, active.

The relation between the stellar mass $M_{\star}$ and the gas mass in a galaxy is adopted from the scaling relations presented in \cite{Tassis_2008}, which are specifically derived for dwarf galaxies without supernova-driven winds. In particular, we use their fiducial model, which suggests, for example, a gas mass of $\sim 10^9\Msun$ for a dwarf galaxy of $\sim 5\times 10^7\Msun$ in stars.
The relation between the gas mass and the gas number density at the Bondi radius is found assuming an isothermal density profile (Eq. \ref{eq:isothermal}). In general, assuming a simple power-law density distribution $\rho(r) = \rho_0(r/r_0)^{-\alpha}$, the gas mass within an effective radius $R_e$ is $M(r<R_e) = 4\pi \rho_0 r_0^{\alpha}R_e^{3-\alpha}/(3-\alpha)$, or $M(r<R_e) = 4 \pi \rho_0 r_0^2 R_e$ in the isothermal case $\alpha = 2$ ($r_0$ is the core radius). With simple operations, the gas number density is
\begin{equation}
    n_{gas} = \frac{M_{gas}}{4 \pi \mu m_h r_0^2 R_e} \, ,
\end{equation}
with $\mu m_h$ the typical mass of molecules in the gas ($\mu$ is the mean molecular mass, $m_h$ is the atomic mass of hydrogen).

We consider effective radii in the range $ 0.01 \lesssim R_e \, [\mathrm{kpc}] \lesssim 5$, based on simulations and observations from \cite{Carraro_2001} and \cite{Forbes_2017}. The first study finds typical effective radii in the range $0.07-1.55$ kpc for dwarfs in the mass range $10^7-10^8 \Msun$, while the second study focuses on more massive galaxies and, in the range $10^9-10^{10} \Msun$, still finds similar values between $0.15$ and $6.30$ kpc. We consider core radii equal to the Bondi radius of the central BH, which scales as $R_B \propto \Mblack/c_s^2$, hence, it also depends strongly on the sound speed of the gas near the BH.
As discussed above we consider here BHs with masses in the range $10^3\Msun$ and $10^7\Msun$ (see \citealt{Bellovary_2019, Greene_review_2020}). The best match between our model and data on the X-ray detected fraction (see \S \ref{sec:results}) is obtained with typical Bondi radii between $0.1$ and $10 \, \mathrm{pc}$, all values significantly smaller than the corresponding effective radii (see also the classic derivation of the sphere of influence of an MBH in \citealt{Bahcall_Wolf_1976}). A variation of two orders of magnitude in the Bondi radius, associated with a variation of four orders of magnitude in the BH mass, implies that the corresponding sound speeds differ by one order of magnitude at most.

The angular momentum parameter $\lambda_{\rm B}$ plays an important role in this model (see Eq. \ref{eq:cond2}), but again cannot be determined directly from observations.
Hence, we express it in terms of quantities that can be estimated observationally.
We define the parameter of rotational support ${\cal R}(r) = v(r)/\sigma(r)$, where $v(r)$ is the rotational velocity of gas/stars measured at a radius $r$ and $\sigma(r)$ is the velocity dispersion at the same location \citep{Bender_1993}. For instance, a value of ${\cal R} > 1$ suggests that the orbits of gas/stars are rotationally supported.
In principle, the parameter ${\cal R}$ should be measured at the Bondi radius $R_B$, i.e. ${\cal R}_B \equiv {\cal R}(r = R_B)$. In practice, the angular resolution of IFS measurements allows an estimate of this parameter only at larger spatial scales (see \S \ref{sec:results}).

We now show how to express $\lambda_{\rm B}$ in terms of our parameter of rotational support ${\cal R}$. We assume an isothermal density profile
\begin{equation}
    \rho(r) = \frac{\sigma^2}{2\pi G r^2} \, ,
    \label{eq:isothermal}
\end{equation}
and recall the definition of $\lambda_{\rm B}$, i.e. $\lambda_{\rm B} = \ell_{\rm B}/(GMR_{\rm B})^{1/2}$, where $\vec{\ell_{\rm B}} = \vec{R_{\rm B}} \times \vec{v}$ ($\times$ indicates the vector product between the two vectors), to obtain 
\begin{equation}
    \lambda_{\rm B} = R_{\rm B} v \sin\theta/(GMR_{\rm B})^{1/2} \, .
\end{equation}
Here, $\theta$ is the deflection angle of the gas at the Bondi radius, or the angle between the direction of motion of the gas and the line connecting the gas to the central point of mass. 
With the expression of the isothermal profile, it is easy to show that the parameter of rotational support, $\cal R$, measured at $R_B,$ can be expressed in terms of the angular momentum parameter, $\lambda_{\rm B}$, as:
\begin{equation}
   {\cal R}_B = \sqrt{\frac{2}{3}} \frac{\lambda_{\rm B}}{\sin \theta} \, ,
   \label{eq:par_rot}
\end{equation}
In the following we use the notation ${\cal R}$ to indicate the parameter measured as close as technically possible to the Bondi radius.
It is interesting to remark that in the case of circular orbit (i.e., $\sin \theta = 1$) we obtain $\lambda_{\rm B} \sim {\cal R}_B$. Also, note that in the case of radial inflow $\theta = 0$, $\sin \theta = 0$, and also $\lambda_{\rm B} = 0$, because there is no angular momentum. Note that ${\cal R}_B/\lambda_{\rm B} = \sqrt{2/3}/\sin{\theta}$ is undefined for $\theta = 0$, as the two-sided limit does not exist.

\subsection{Calculating the active fraction}
\label{subsec:act_frac}
To estimate the active fraction, ${\cal A}$, in a dwarf galaxy of stellar mass $M_{\star}$ we adopt a probabilistic approach, and ask: what is the BH mass range that would allow a BH to be active in a given galaxy and how does it compare with the full BH mass range considered in dwarfs (i.e., $10^3 - 10^7 \Msun$)? The ratio of these two mass ranges is used as an estimator of the probability that the BH is active.

There is a fundamental assumption in this approach: we assume that in a dwarf galaxy the occupation probability for a BH in the mass range $10^3 - 10^7 \Msun$ is uniform (in the logarithm of the mass). In other words, the prior for the BH occupation fraction in dwarfs is flat. This assumption is powerful as it is not based on the validity of the $\Mblack-\sigma$ relation in dwarfs, as some studies have suggested a departure from it at lower masses (e.g., \citealt{Martin-Navarro_2018, Pacucci_2018, Nguyen_2019}), while others (e.g., \citealt{Baldassare_2020}) hint that it might hold. Consequently, \textit{all the formulas below involving BH masses are to be intended as expressed in the logarithm in base 10, unless otherwise stated}.

For a given galaxy of stellar mass $M_{\star}$, we define $M_{\bullet, act}$ as the range of BH masses that allows an efficient accretion. Figure \ref{fig:galaxy_BH} shows an example of mass range $M_{\bullet, act}$ for a galaxy with central gas number density $\sim 10^5 \, \mathrm{cm^{-3}}$, indicated with a green, hatched region. This value is calculated as: $M_{\bullet, act} = M_{\bullet, max} - \max(M_{\bullet, S}, M_{\bullet, L})$, where $M_{\bullet, max} = 7$ and, similarly, $M_{\bullet, min} = 3$, with $\Delta M = M_{\bullet, max} - M_{\bullet, min}$ (shown in a gray, hatched region in Fig. \ref{fig:galaxy_BH}). The active fraction is then computed as:
\begin{equation}
    {\cal A} = M_{\bullet, act}/\Delta M \, .
\end{equation} 

As the mass range $M_{\bullet, act}$ increases with the stellar mass of the host, more massive galaxies are predicted to have a higher active fraction.

\section{Model calibration} 
\label{sec:calibration}
We need an intermediate step to predict the multiwavelength active fraction for dwarf galaxies and estimate, en passant, the value of some physical properties of the innermost volume of the hosts. Although active fractions are bounded from above by the occupation fraction, they are not easily determined by observations, and some extensive studies based on data are only available in the X-ray domain. Hence, we first calibrate our model to the observationally constrained X-ray detected active fraction (e.g., \citealt{Mezcua_2018, Birchall_2020}). 

All active BHs predicted in our model have X-ray emission, but not all of them can be detected in an X-ray survey. In order to compare our predicted active fraction with the X-ray detected fraction, we consider the following two conditions that can limit our ability to detect, in X-rays, actively accreting BHs in dwarfs:
\begin{enumerate}
    \item AGN with an X-ray luminosity lower than the X-ray emission from star-forming regions in the galaxy are not identified as such. In addition, AGN with X-ray luminosity fainter than $\sim 10^{39} \, \mathrm{erg \, s^{-1}}$ are challenging to detect \citep{Mezcua_2016, Mezcua_2018, Birchall_2020}.
    \item AGN buried inside Compton-thick column densities ($N_H \gtrsim 10^{24} \, \mathrm{cm^{-1}}$) may be undetectable in the X-rays (see, e.g., \citealt{Hickox_2018}).
\end{enumerate}

Regarding (1), an estimate of the X-ray emission from star-forming regions and of the X-ray emission from an MBH, both as a function of the stellar mass of the dwarf, was derived from \cite{Aird_2017} and \cite{Birchall_2020}, respectively.
The typical X-ray emission from star-forming regions of galaxies was estimated with the following equation derived in \cite{Aird_2017}:
\begin{equation}
\begin{split}
    \mathrm{log_{10}L_x} = & a + b \left( \mathrm{log_{10}} \frac{M_{\star}}{\Msun} -10.2 \right) + \\
    & + c \, \mathrm{log_{10}} \left( \frac{1+z}{2} \right) \, .
    \label{eq:aird}
\end{split}
\end{equation}
Here, $a = 40.76 \pm 0.02$, $b = 0.63 \pm 0.03$ and $c = 3.79 \pm 0.12$, $M_{\star}$ is the stellar mass of the galaxy. We consider $z\sim 0.1$ as the typical redshift of our sample.
For an MBH to be detected we require its X-ray emission to be brighter (at a confidence level of $3\sigma$) than the emission from star formation. The confidence level is estimated from the values of the uncertainties in Eq. \ref{eq:aird}. Note that a fraction of dwarf galaxies are characterized by low metallicity (see, e.g., \citealt{Kirby_2013}), which makes the X-ray detection of slowly accreting MBHs even more challenging. In fact, in low-metallicity environments the X-ray emission from star-forming regions is enhanced \citep{Fragos_2013, Brorby_2016, Fornasini_2020}. This is coupled to an apparent deficit in X-ray luminosity from BH accretion \citep{Simmonds_2016, Cann_2020}. We discuss the effect of low-metallicity environments in Sec. \S \ref{sec:results}.
Regarding (2), an MBH might be actively accreting but its X-ray emission could be undetected because of heavy obscuration. Several studies (see, e.g., \citealt{Comastri_2015, Ananna_2019}) and reviews (e.g., \citealt{Hickox_2018}) suggest that a considerable fraction of AGN may be heavily obscured. For this study we employ the recent estimate in \cite{Ananna_2019}, which finds that $\sim 56 \pm 7 \%$ of AGN within $z\lesssim 1$ are Compton thick.

The factors (1) and (2), to correct for the issues described above, are used to transform from the active fraction to the X-ray detected fraction. Hence, we first calibrate our model parameters against the X-ray detected fraction data, then we remove the normalizing parameters described in points (1) and (2) to derive the multiwavelength active fraction.
We are aware that these numerical corrections are subject to large uncertainties. However, also the X-ray detected fractions in \cite{Birchall_2020, Mezcua_2018} are associated with uncertainties as large as $\sim 35\%$, so our calibration to the X-ray detected fraction is to be considered an order-of-magnitude estimate.

Calibration is needed to estimate the angular momentum content of the innermost volume of the hosts as a function of their stellar mass. The angular momentum of the gas in the inner regions of the dwarf galaxy, quantified with the parameter $\lambda_{\rm B}$, is a fundamental parameter in our model but cannot be probed observationally. 
We search for suitable expressions of the function $\lambda_{\rm B} = \lambda_{\rm B}(M_{\star})$, and find that a simple power-law relation of the form
\begin{equation}
    \lambda_{\rm B} = \alpha \, (\log_{10}{M_{\star}})^{\beta} \, ,
    \label{eq:lambda_to_Mstar}
\end{equation}
works well. In this expression, the only known information is that the stellar mass considered varies within $\log_{10}M_{\star, min} = 5$ and $\log_{10}M_{\star, max} = 10$ (see, e.g., \citealt{Bellovary_2019}). As we do not know the range of variation for $\lambda_{\rm B}$, we perform an automated search of the parameter space, in the range $10^{-6} < \lambda_{\rm B} < 10^{-2}$, until a reasonable agreement with the X-ray data is found. The values of the parameter $\lambda_{\rm B}$ corresponding to the maximum and minimum mass are found to be $\lambda_{\rm B, max} = 2.1\times 10^{-4}$ and $\lambda_{\rm B, min} = 4.2\times 10^{-4}$.
Note that $\beta$ can then be expressed as:
\begin{equation}
    \beta = \log_{2} \, \left( \frac{\lambda_{\rm B, max}}{\lambda_{\rm B, min}} \right) \, .
    \label{eq:beta}
\end{equation}
The base of the logarithm in the previous expression comes from $\log_{10}M_{\star, max}/\log_{10}M_{\star, min}$.
In turn, $\alpha$ is solved as $\alpha = \lambda_{\rm B, max} / (\log_{10}M_{\star, max})^b$. 
The reader should note that this is not a typical regression because the values of the parameter $\lambda_{\rm B}$ are unknown. Hence, we proceed in this way: (i) we iteratively choose couples of $\lambda_{\rm B, max}, \lambda_{\rm B, min}$ within the parameter range mentioned above; (ii) we calculate the corresponding $\alpha$ and $\beta$ and use the relation $\lambda_{\rm B} = \lambda_{\rm B}(M_{\star})$ to compute the X-ray detected active fraction; (iii) we compare it with calibration data and choose, by eye, the combination of parameters that offers a good resemblance to data. Of course, this procedure does not guarantee that the solution found is unique. Nonetheless, we verify that all the combinations of $\lambda_{\rm B}$ that offer a good calibration to data are of order $\sim 5\times 10^{-4}$.
In conclusion, our fiducial choice of parameters is: $\alpha = 0.0021$ and $\beta = -1.0$.

\begin{figure}[h]
\includegraphics[angle=0,width=0.49\textwidth]{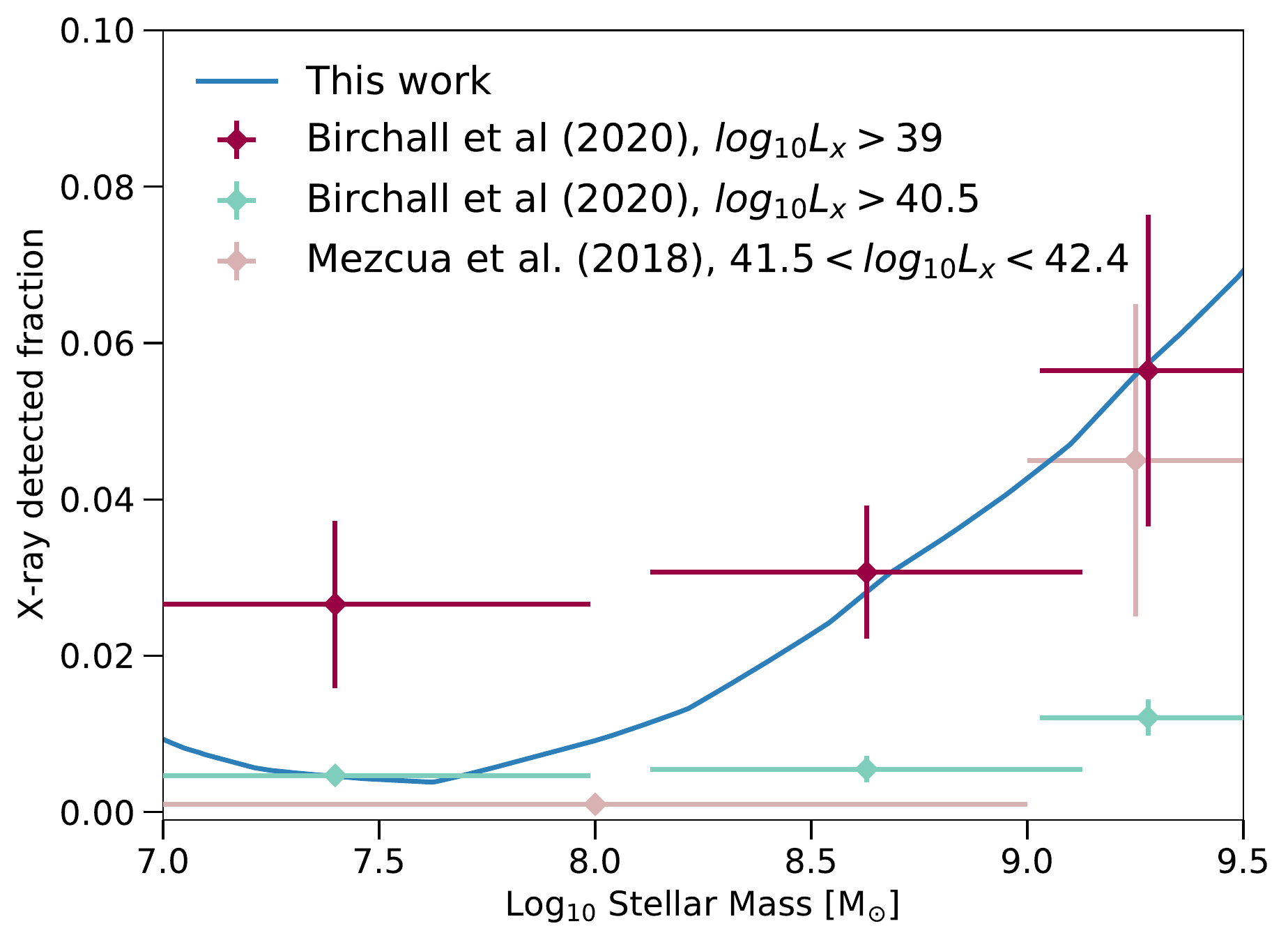}
\caption{The X-ray detected fraction in local ($z < 0.25$) dwarfs as a function of the stellar mass of the host galaxy. Our model for the multiwavelength active fraction is calibrated against these data, for several X-ray luminosity ranges, and corrected to take into account that some active BHs might not be detectable in the X-ray domain. The upturn of the X-ray detected fraction (and active fraction, see Fig. \ref{fig:occ_frac}) at low stellar mass is due to the breaking of some of the assumptions in our model, including the flat prior in the distribution of BH masses. Data are obtained from \cite{Mezcua_2018} and \cite{Birchall_2020}.} 
\label{fig:x-ray-detected}
\end{figure}

\begin{figure*}
\begin{center}
\includegraphics[angle=0,width=0.80\textwidth]{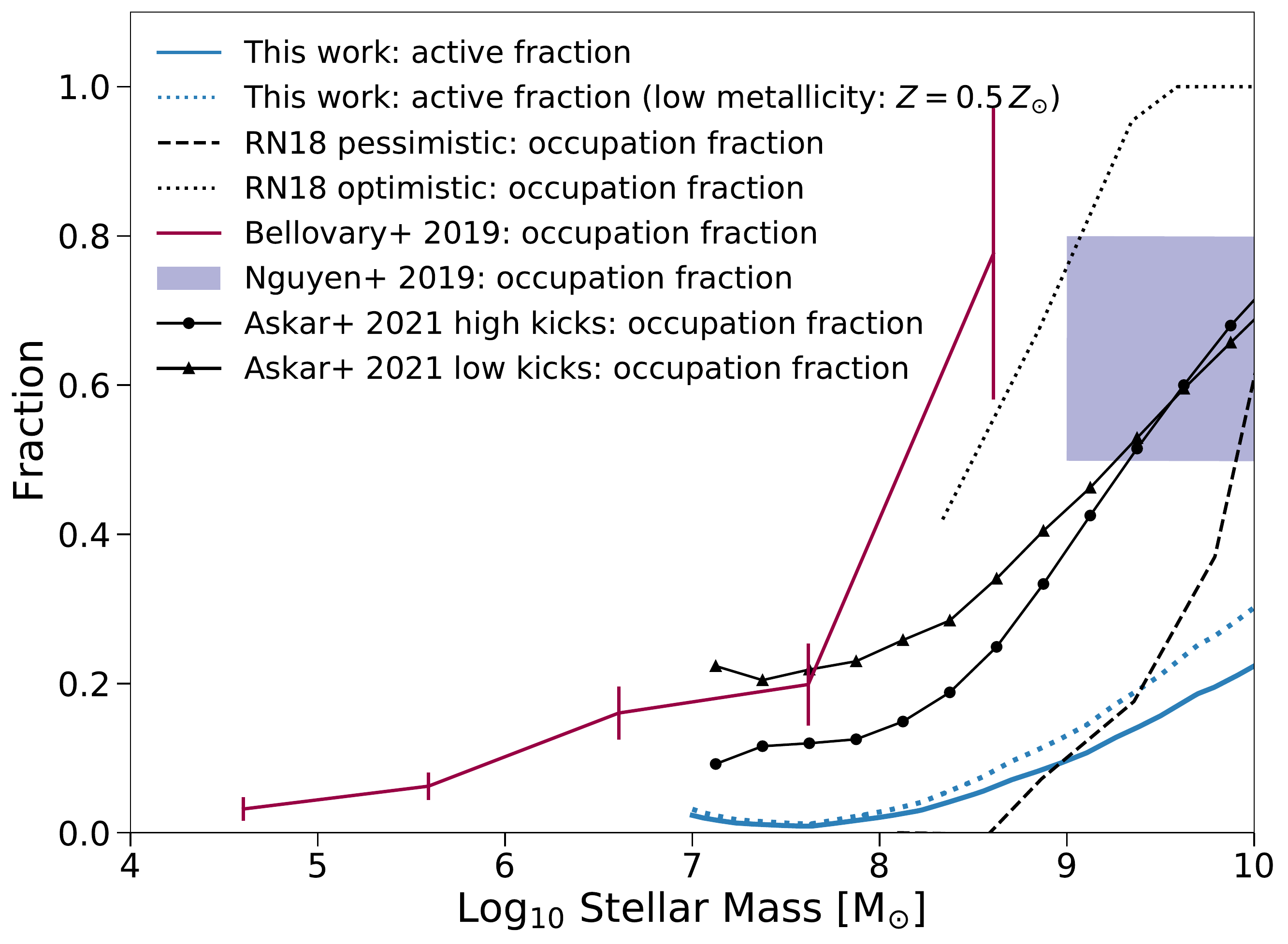}
\caption{Our model for the active fraction (blue solid line) is compared with several \textit{occupation fractions} calculated in \cite{Ricarte_2018}, RN18, \cite{Bellovary_2019}, \cite{Nguyen_2019}, \cite{Askar_2021}. As expected, our active fractions are lower than most of the predictions for occupation fraction. The blue dotted line indicates the active fraction computed assuming that dwarf galaxies have low-metallicity environments with $Z=0.5 \, Z_{\odot}$.}
\end{center}
\label{fig:occ_frac}
\end{figure*}

The computed X-ray detected fraction is shown with X-ray data in Fig. \ref{fig:x-ray-detected}, considering the entire local ($z<0.25$) sample from \cite{Birchall_2020} and the $z<0.2$ subsample in \cite{Mezcua_2018}. For low stellar masses, $\lesssim 10^{8} \Msun$, our calibration tends to indicate X-ray detected fractions of the order of $\sim 1\%$, more similar to the subsamples including only the brightest sources, with $L_\mathrm{x} > 10^{40.5} \, \mathrm{erg \, s^{-1}}$ from \cite{Birchall_2020} and with $ 10^{41.5} < L_\mathrm{x} \, [\mathrm{erg \, s^{-1}}] < 10^{42.4}$ from \cite{Mezcua_2018}. This might be due to the fact that some of the fainter objects included in the subsample by \cite{Birchall_2020} ($L_\mathrm{x} > 10^{39} \, \mathrm{erg \, s^{-1}}$) are interpreted as nondetectable by our model, because their emission is confused with that from star formation (see the beginning of this section).
The slight upturn of the X-ray detected fraction at stellar masses $\lesssim 10^{7.5} \Msun$, visible also in the main results for the active fraction (see Fig. \ref{fig:occ_frac}) is due to the fact that some assumptions in our model are breaking down at BH masses $\lesssim 10^3 \Msun$, including the flat prior in the distribution of BH mass. In fact, at this mass scale we are entering the realm where many BHs of similar mass exist in the same host, so that the occupation fraction is artificially increased. Furthermore, below the $\Mblack = 10^3 \Msun$ scale, we are exiting the limits in which a central MBH is defined and entering the regime of stellar mass BHs that are not expected to be centrally located and should have an occupation fraction of 1. For large stellar masses, $\gtrsim 10^{8.5} \Msun$, the calibration follows the upturn shown, to various degrees, by all the data samples.

\section{RESULTS} 
\label{sec:results}
We are now in a position to estimate the multiwavelength active fraction for MBHs which is, by construction, equal or lower than the occupation fraction for any given stellar mass.

\subsection{The active fraction} 
\label{subsec:fractions}
Taking into account all of the considerations presented in \S \ref{sec:calibration} we can now compute the multi-wavelength active fraction. In other words, subtracting away the limitations of X-ray observations, we are now calculating the actual active fraction of BHs, which could, in principle, be detected in any wavelength. This is then expressed as a function of both the stellar mass and of the parameter of rotational support.

The resulting plot is shown in Fig. \ref{fig:occ_frac}, where our model is compared with several \textit{occupation fractions} from \cite{Ricarte_2018, Bellovary_2019, Nguyen_2019, Askar_2021}. We find multiwavelength active fractions for MBHs ranging from 5\% to 22\%, and increasing with the stellar mass of the host as $\sim (\log_{10}M_{\star})^{4.5}$. Note that this active fraction is for MBHs accreting at typically low Eddington ratios. In accordance with our expectations, our prediction for the active fraction is lower than most models for the occupation fraction, except a small mass range in the model by \cite{Ricarte_2018} with pessimistic assumptions on seeding. Our prediction for the active fraction terminates at a stellar mass $\sim 10^7 \Msun$. At lower stellar masses we would expect BHs that are lighter than $\Mblack = 10^3 \Msun$, which is the lower bound of BH mass considered in this study. As anticipated in \S \ref{sec:calibration}, the upturn at low stellar mass is likely due to the breakdown of some assumptions in our model, in particular, the presence of several BHs of similar mass in the same host. Hence, our flat prior on the BH mass ceases to hold because very light BHs have an intrinsically higher probability to be found inside a galaxy than an MBH. 

In Sec. \S \ref{sec:calibration} we anticipated that the X-ray detection of MBHs in low-metallicity dwarf galaxies can present enhanced challenges, due to increased X-ray luminosity from star-forming regions \citep{Fragos_2013, Brorby_2016, Fornasini_2020} and reduced X-ray luminosity from BH accretion \citep{Simmonds_2016, Cann_2020}. These two effects both increase the correction factor (1) described in Sec. \S \ref{sec:calibration}, i.e., current X-ray surveys are able to detect a smaller fraction of the BHs that are actively accreting. Hence, \textit{low-metallicity effects would increase our predicted active fraction} shown in Fig. \ref{fig:occ_frac}. As an example, we discuss the case of a decrease in metallicity from $Z = Z_{\odot}$ to $Z = 0.5 \, Z_{\odot}$, where $Z_{\odot}$ is the solar metallicity. Following \cite{Fornasini_2020}, and assuming that the star formation rate of the galaxy is constant, we obtain that $Z = Z_{\odot}$, or $12 + \mathrm{log_{10}}(\mathrm{O/H}) = 8.69$, corresponds to a typical X-ray luminosity from X-ray binaries of $10^{39.4} \, \mathrm{erg \, s^{-1}}$, while $Z = 0.5 \, Z_{\odot}$, $12 + \mathrm{log_{10}}(\mathrm{O/H}) = 8.39$, corresponds to $10^{39.7} \, \mathrm{erg \, s^{-1}}$, i.e. a factor $\sim 2$ increase. Regarding the decreased X-ray luminosity from BH accretion, recent studies \citep{Simmonds_2016, Cann_2020} are detecting sources in low-metallicity dwarf galaxies showing broad optical emission lines, typical of AGN activity, but X-ray luminosity $\sim 1-2$ orders of magnitude lower than expected. Assuming a conservative value of one order-of-magnitude decrease and a factor $2$ increase in the X-ray luminosity from star-forming regions, we estimate that current X-ray surveys may be missing $\sim 35\%$ of actively accreting MBHs in the mass range considered. This results in a slightly enhanced active fraction prediction for dwarfs, shown as a dotted blue line in Fig. \ref{fig:occ_frac}. This may reach $\sim 30\%$ for the most massive hosts.

Figure \ref{fig:2D} shows a contour plot for the active fraction calculated as a function of the stellar mass and of the angular momentum parameter $\lambda_{\rm B}$ (right axis). The values of the function $\lambda_{\rm B} = \alpha \, (\log_{10}{M_{\star}})^{\beta}$ constrained by the X-ray detected fractions are shown with a white dashed line ($\alpha = 0.0021$ and $\beta = -1.0$, see Eq. \ref{eq:lambda_to_Mstar}). The values of the parameter $\lambda_{\rm B}$ required to fit the X-ray data are of the order $\lambda_{\rm B}$ $\lesssim 10^{-3}$. As expected, a lower value of $\lambda_{\rm B}$ leads to a larger value of the active fraction, as the rotational support of the gas is diminished.

In the following section, we show how these values can be reconciled with values of the parameter of rotational support inferred from IFS observations of AGN in nearby dwarf galaxies. This will require the use of an additional free parameter of the model: the angle of incidence of the gas at the Bondi radius, $\theta$ (see Eq. \ref{eq:par_rot}).

\subsection{Comparing $\lambda_{\rm B}$ with ${\cal R}$ from IFS observations} 
\label{subsec:constraining_rot_sup}
In Eq. \ref{eq:par_rot} we showed that, in an isothermal density profile, $\lambda_{\rm B}$ can be expressed in terms of the parameter of rotational support at the Bondi radius, ${\cal R}_{\rm B} = v_{\rm B}/\sigma_{\rm B}$, where $v_{\rm B}$ and $\sigma_{\rm B}$ are the rotational velocity and the velocity dispersion of stars/gas, respectively. 
The rationale for this choice is that $\lambda_{\rm B}$ is very challenging to measure from observations, while ${\cal R}$ can be estimated with IFS. We recognize that this technique cannot directly probe the Bondi radii of MBHs studied here, but at least can provide an order-of-magnitude estimate of the quantities involved.

We use recent IFS observations of AGN in dwarf galaxies to derive the rotational velocity and velocity dispersion of the [OIII] gas and of the stars at the center of these galaxies. The data are collected as a part of the study of \cite{Mezcua_2020_rot_sup}, who find 37 AGN in dwarf galaxies based on emission line diagnostics derived from IFS data of the SDSS/MaNGA (Mapping Nearby Galaxies at APO; \citealt{Bundy2015}) survey. These AGN are local, with typical redshift $z \sim 0.02$. We select those dwarf galaxies (14 out of 37) with central AGN (i.e. those with an AGN also according to the emission line diagnostics derived from the central SDSS single-fiber aperture), and derive the [OIII] and stellar velocities at the central pixel ($0.5''$) of the MaNGA IFS kinematic maps. The data used in this study are reported in Table \ref{table1} of the Appendix \ref{Appendix}. The data suggest that both the gas and the stars at the center of these galaxies are characterized by parameters of rotational support in the range ${\cal R} \sim 10^{-1} - 10^{-2}$. Although this measurement probes only the central $0.5''$ pixel, i.e. on scales much larger than the Bondi radius, we consider this as an indication of what a typical range for this parameter looks like: ${\cal R} \gtrsim 10^{-2}$.

For one of the galaxies in the Appendix with a clear rotational velocity profile (MaNGA plateifu $7958-9101$, see Table \ref{table1}) we perform a power-law regression of the value of ${\cal R}$ as a function of the angular distance from the galactic centroid $\delta$, in order to extrapolate the value at the Bondi radius of the MBH (Fig. \ref{fig:7958-9101} in the Appendix, bottom right panel). We obtain the best-fit relation:
\begin{equation}
    {\cal R}(\delta) = (0.695 \pm 0.005) \, \delta^{(0.669 \pm 0.006)} \, ,
    \label{eq:extrapolation}
\end{equation}
where the separation $\delta$ is expressed in arcseconds.
The galaxy $7958-9101$ is at $z=0.039$ and with a typical Bondi radius of $\sim 1$ pc we would need to probe scales of the order of $0.001''$. Extrapolating the value of ${\cal R}$ by using Eq. \ref{eq:extrapolation} with $\delta=0.001$, we obtain ${\cal R} \sim 7 \times 10^{-3}$. Hence, we conclude that a better estimate of the parameter of rotational support at scales comparable to the Bondi radii is ${\cal R} \gtrsim 5\times 10^{-3}$. The other three panels of Fig. \ref{fig:7958-9101} show the kinematic maps of the [OIII] gas, specifically its velocity, velocity dispersion, and rotational support parameter ${\cal R}$.

\begin{figure*}
\begin{center}
\includegraphics[angle=0,width=0.80\textwidth]{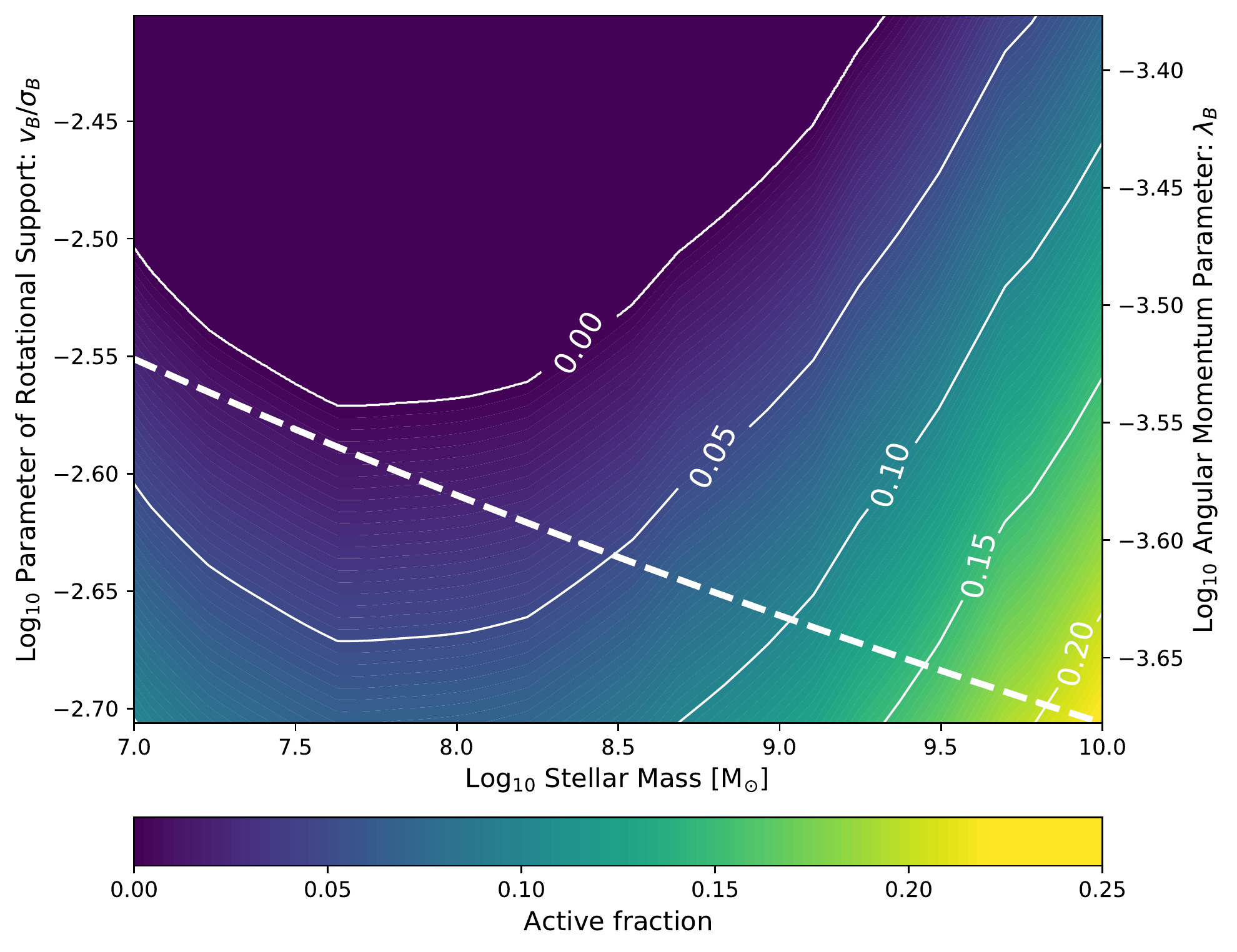}
\caption{Predicted active fraction shown as a function of $M_{\star}$, in terms of ${\cal R}$ on the \textbf{left axis} and in terms of $\lambda_{\rm B}$ on the \textbf{right axis}. The parameter of rotational support ${\cal R}$ is in the range estimated from observations by assuming a value of the angle $\theta \sim 5$ degrees. The dashed lines indicate the preferred relation between $M_{\star}$ and (${\cal R}$, $\lambda_{\rm B}$), calculated by calibrating our model against the X-ray detected fraction. The solid white lines are the contour levels for the active fraction. Note that the left axis is obtained by rescaling the right axis by a factor $\sqrt{2/3}/\sin\theta$, with $\theta = 5$ degrees.}
\end{center}
\label{fig:2D}
\end{figure*}

For a circular orbit of the gas at the Bondi radius ($\theta=90$ degrees) we obtain from Eq. \ref{eq:par_rot}: $\lambda_{\rm B} \sim {\cal R}$. As $\lambda_{\rm B} \sim 5\times 10^{-4}$ (see Fig. \ref{fig:2D}), we propose some possible solutions to the discrepancy of $\sim 1$ order of magnitude between our predicted values of $\lambda_{\rm B}$ and the measured values of ${\cal R}$:
\begin{enumerate}
    \item The IFS observations are not able to probe the regions close to the Bondi radius of the BH, and the effective value of ${\cal R}$ is dropping off with decreasing radius faster than what we can probe by extrapolation. Its measured value is, thus, overestimated (i.e., the gas/stars are less rotationally supported at smaller scales than what measured). If this is the case, our model would work with almost circular orbits: $\theta \sim 90$ degrees.
    \item The IFS observations are, on the contrary, accurate in estimating the parameter of rotational support, hence our model would reconcile with these IFS observations with a smaller angle $\theta \lesssim 10$ degrees.
    \item More likely, a combination of the two points above is the correct solution.
\end{enumerate}
Note that the value of $\theta$ is important because, ultimately, it discriminates between a circular vs. radial accretion on to the central MBH.

In Fig. \ref{fig:2D} we show the active fraction predicted in our model as a function of the stellar mass of the dwarf and of the parameter of rotational support (left axis), assuming $\theta = 5$ degrees. Note that the left axis is obtained just rescaling the right axis by a factor $\sqrt{2/3}/\sin(5)$. The active fraction can be interpreted as the probability of hosting an active MBH, accreting at typically low Eddington ratios.

The dashed white line indicates the preferred relation between $M_{\star}$ and ${\cal R}$:
\begin{equation}
    {\cal R}(M_{\star}) = \sqrt{\frac{2}{3}} \, \frac{\alpha \, (\log_{10}{M_{\star}})^{\beta}}{\sin(\theta)} \, .
    \label{eq:R_M}
\end{equation}
which is found by eliminating $\lambda_{\rm B}$ between Eqs. \ref{eq:par_rot} and \ref{eq:lambda_to_Mstar} and by choosing $\alpha = 0.0021$, $\beta = -1.0$ and $\theta = 5$ degrees.

\begin{figure}
\includegraphics[angle=0,width=0.49\textwidth]{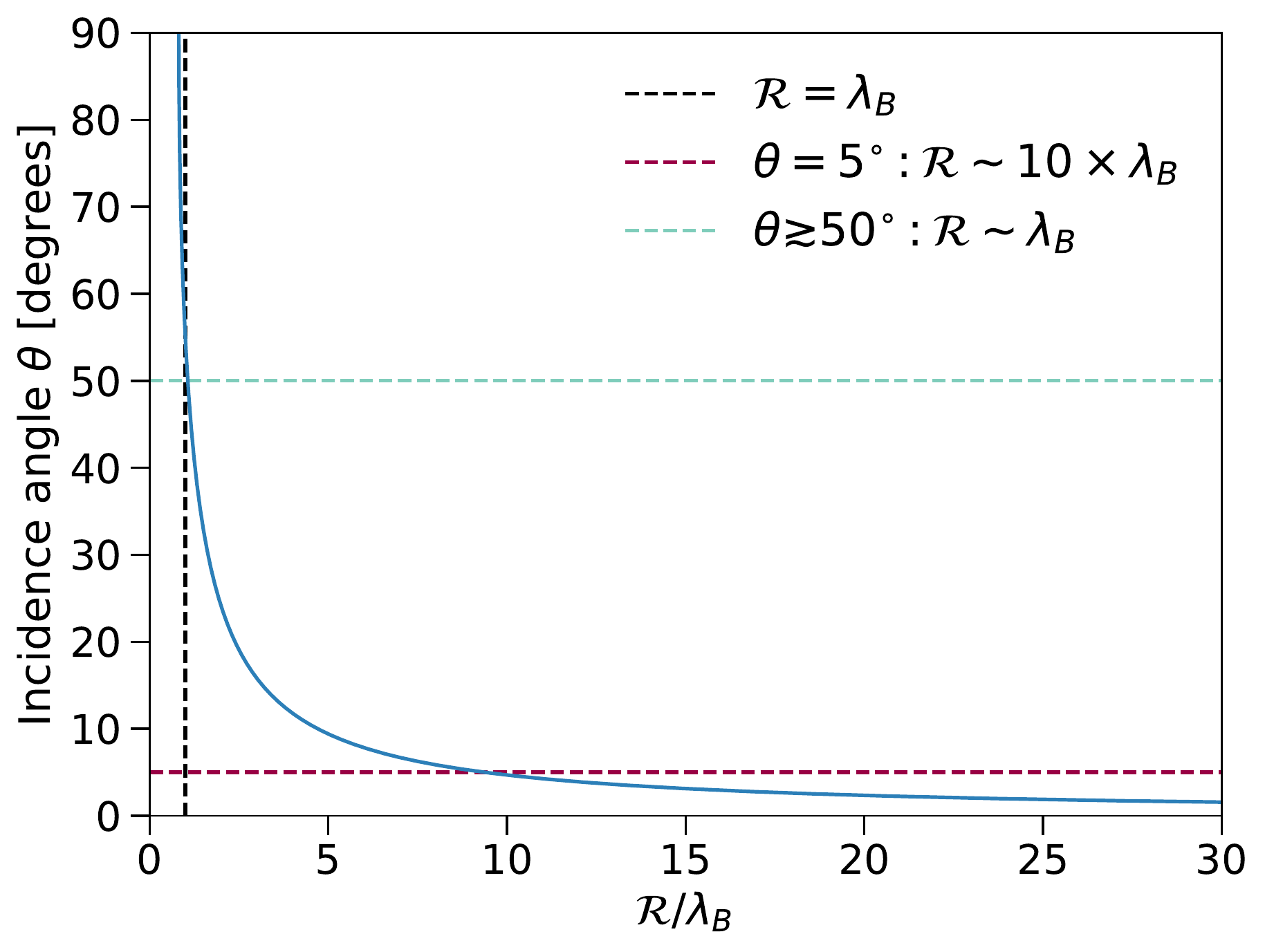}
\caption{The incidence angle $\theta$ (in degrees) required to explain a given ratio between the observationally determined ${\cal R}$ and the predicted $\lambda_{\rm B}$. The black dashed vertical line indicates $\lambda_{\rm B} = {\cal R}$. The red dashed line indicates the angle of $5$ degrees required by the current IFS data. The green dashed line shows that for any value $\theta \gtrsim 50$ degrees $\lambda_{\rm B} \sim {\cal R}$.}
\label{fig:theta}
\end{figure}

In Fig. \ref{fig:theta} we show the incidence angle $\theta$ required to explain a given ratio between the observationally determined ${\cal R}$ and the predicted $\lambda_{\rm B}$. The red dashed line indicates the angle of $5$ degrees required by the current IFS data. The green dashed line shows that the model is insensitive to values of $\theta \gtrsim 50$ degrees when $\lambda_{\rm B} \sim {\cal R}$.

For convenience, the nonzero region of the phase plot in Fig. \ref{fig:2D} for ${\cal A}$ can be approximated in terms of $M_{\star}$ and ${\cal R}$ with a simple quadratic surface without mixed terms
\begin{multline}
    {\cal A}(M_{\star},{\cal R}) \approx 0.54490 -0.52198 M_{\star} + \\ 
    +0.03341 M_{\star}^2 -0.66946 {\cal R} -0.03373 {\cal R}^2 \, .
    \label{eq:fit}
\end{multline}
The quadratic relation ${\cal A}(M_{\star},{\cal R})$ shows a higher active fraction for MBHs in a dwarf galaxy with a larger stellar mass and a lower value of the parameter of rotational support in the innermost regions of the galaxy, as expected.

Several studies have investigated the rotational support in galaxies, both using observations of dwarfs (e.g., \citealt{Wheeler_2017}) and simulations (e.g., \citealt{ElBadry_2018}). In particular, \cite{Wheeler_2017} find that the stars in $\sim 80\%$ of the local dwarf galaxies studied are dispersion supported. Additionally, they find that $\sim 60\%$ of isolated dwarfs in the sample have $v/\sigma < 1$. The values they find refer to the galaxies as a whole and are larger by orders of magnitude with respect to the ${\cal R} \sim 10^{-1} - 10^{-2}$ we possibly find at the center of dwarfs. 
The decrease of the ratio $v/\sigma$ with decreasing distance from the center of the dwarf is expected, as studies (e.g., \citealt{Rhee_2004, Regan_2014}) have shown that, also in dwarfs, the circular velocity decreases, while the velocity dispersion increases toward the center.

\section{Discussion and Conclusions} 
\label{sec:disc_concl}
Since the beginning of the XXI century we have known that most massive galaxies host a SMBH at their center, and that a profound correlation exists between its mass and some physical properties of the galaxy (e.g., \citealt{Ferrarese_Merritt_2000, Gebhardt_2000}). More recently we have started to investigate these relationships for lighter BHs, namely, intermediate-mass BHs, typically defined in the mass range $10^3 \lesssim \Mblack (\Msun) \lesssim 10^{6}$ (e.g., \citealt{Mezcua_2017_review}; \citealt{Greene_review_2020}). Furthermore, the occupation fraction and the active (or AGN) fraction of galaxies, at least in the X-ray domain, are subject in recent years of intense focus from the scientific community, with studies approaching the problem with cosmological simulations, SAMs and observations.

In this study, we presented a physical model to predict the multiwavelength (i.e., not restricted to any frequency range) active fraction, ${\cal A}$, of MBHs in dwarf galaxies.
This model aims to predict the active fraction starting from first principles, using as inputs some basic physical properties of the galaxies, such as their gas density and angular momentum content. These properties are then expressed in terms of observational proxies that can be more easily measured.
We calibrate our model against X-ray detected AGN fractions, and calculate the multiwavelength active fraction for dwarf galaxies, with a stellar mass in the range $10^7-10^{10} \Msun$. 
The active fraction can be interpreted as the probability that a dwarf galaxy hosts an actively accreting MBH and it is lower, by construction, than the occupation fraction. 

The angular momentum content of the innermost volume of the galaxy is challenging to measure observationally, thus we used as proxy the parameter of rotational support \citep{Bender_1993} ${\cal R} = v/\sigma$. This parameter indicates whether the motion of gas/stars is rotationally supported at a given location and was expressed in terms of the logarithm of the stellar mass of the host as ${\cal R}(M_{\star}) = \sqrt{2/3} \, \alpha \, (\log_{10}{M_{\star}})^{\beta}/\sin(\theta)$, with $\alpha = 0.0021$ and $\beta = -1.0$.
Observations of local AGN with IFS techniques suggest a value of ${\cal R} \gtrsim 5\times 10^{-3}$, which can be reconciled with our predictions assuming an angle of incidence $\theta \lesssim 10$ degrees. Probing smaller volumes at the center of the host galaxy will be crucial to have a better estimate of the angle $\theta$.

The three main results of this study can be summarized as follows:
\begin{itemize}
    \item Multiwavelength active fractions for MBHs in dwarf galaxies, accreting at typically low rates, ranging from 5\% to 22\%. In dwarf galaxies with low-metallicity environments the active fraction may reach $\sim 30\%$ in the most massive hosts.
    \item The active fraction increases with the stellar mass of the host as ${\cal A} \sim (\log_{10} M_{\star})^{4.5}$ for galaxies with stellar mass in the range $10^7<M_{\star} [\Msun]<10^{10}$.
    \item We find a fitting formula to predict the active fraction ${\cal A}$ in a dwarf galaxy from observationally derived data on stellar mass and parameter of rotational support ${\cal R} = v/\sigma$ calculated in the inner volume of the hosts: \\${\cal A}(M_{\star},{\cal R}) \approx 0.54490 -0.52198 M_{\star} + 0.03341 M_{\star}^2 -0.66946 {\cal R} -0.03373 {\cal R}^2$.\\
\end{itemize}

\noindent Detecting MBHs in dwarfs is a challenging task, due to their overall low luminosities, but also a fundamental one. MBHs are the link between the two well-known populations of BHs, stellar-mass, and supermassive ones. A better understanding of their demography will be beneficial for a vast number of studies, encompassing, e.g., galaxy-BH coevolution (e.g., \citealt{Greene_review_2020, Baldassare_2020}) and early BH seed formation (e.g., \citealt{Haiman_Loeb_2001, Pacucci_2018}). Our model can be instrumental to assess the probability of finding MBHs in a given galaxy and to better direct observational efforts in the coming years.

The upcoming JWST mission will play a central role in detecting faint AGN in dwarf galaxies, which we are blind to in current surveys, due to both their inherent faintness and due to (X-ray) obscuration. JWST will offer the capability to expand our catalog of known faint AGN both by direct observations and by dynamical measurements.
Recent studies (e.g., \citealt{Cann_2018, Satyapal_2021}) have pointed out the importance of near-infrared and mid-infrared diagnostics for the detection of slowly accreting MBHs. Common techniques based on radio, infrared, optical, and X-ray diagnostics are subject to false negatives due to the presence of intense star formation and/or obscuration in the host galaxy (e.g., \citealt{Trump_2015, Satyapal_2021}). These studies suggested the use of diagnostics based on near-infrared and mid-infrared emission lines to detect elusive MBHs, perfectly suited to the NIRSpec and MIRI spectrometers onboard JWST. These new techniques could detect MBHs with typical Eddington ratios $f_{\rm Edd} \sim 0.1$ (in accordance with our assumptions) in the local Universe and up to $z\sim 3$. If these predictions prove reliable, JWST could detect active fractions possibly a factor $\sim 3$ larger than current X-ray surveys \cite{Satyapal_2021}.

In summary, MBHs in dwarfs are elusive, but extremely important for our understanding of the whole population of BHs, spanning $\sim 10$ orders of magnitude in mass. Future theoretical and observational efforts will hopefully shine more light on this missing chain in the BH population.

\vspace{0.3cm}

\section*{Acknowledgments}
We thank the anonymous referee for constructive comments on the manuscript.
F.P. acknowledges fruitful discussions with Avi Loeb and Ramesh Narayan, and support from a Clay Fellowship administered by the Smithsonian Astrophysical Observatory. This work was supported in part by the Black Hole Initiative at Harvard University, which is funded by grants from the John Templeton Foundation and the Gordon and Betty Moore Foundation. This work was also partly performed at the Aspen Center for Physics, which is supported by National Science Foundation grant PHY-1607611. The participation of F.P. at the Aspen Center for Physics was supported by the Simons Foundation. M.M. acknowledges support from the Beatriu de Pinos fellowship (2017-BP-00114) and the Ramon y Cajal fellowship (RYC2019-027670-I). J.R. acknowledges support from the Royal Society and Science Foundation Ireland under grant number URF$\backslash$R1$\backslash$191132.





\bibliographystyle{mnras}
\bibliography{ms}

\clearpage
\appendix

\section{Parameter of rotational support from IFS data}
\label{Appendix}
\begin{figure*}
\centering
\includegraphics[width=\textwidth]{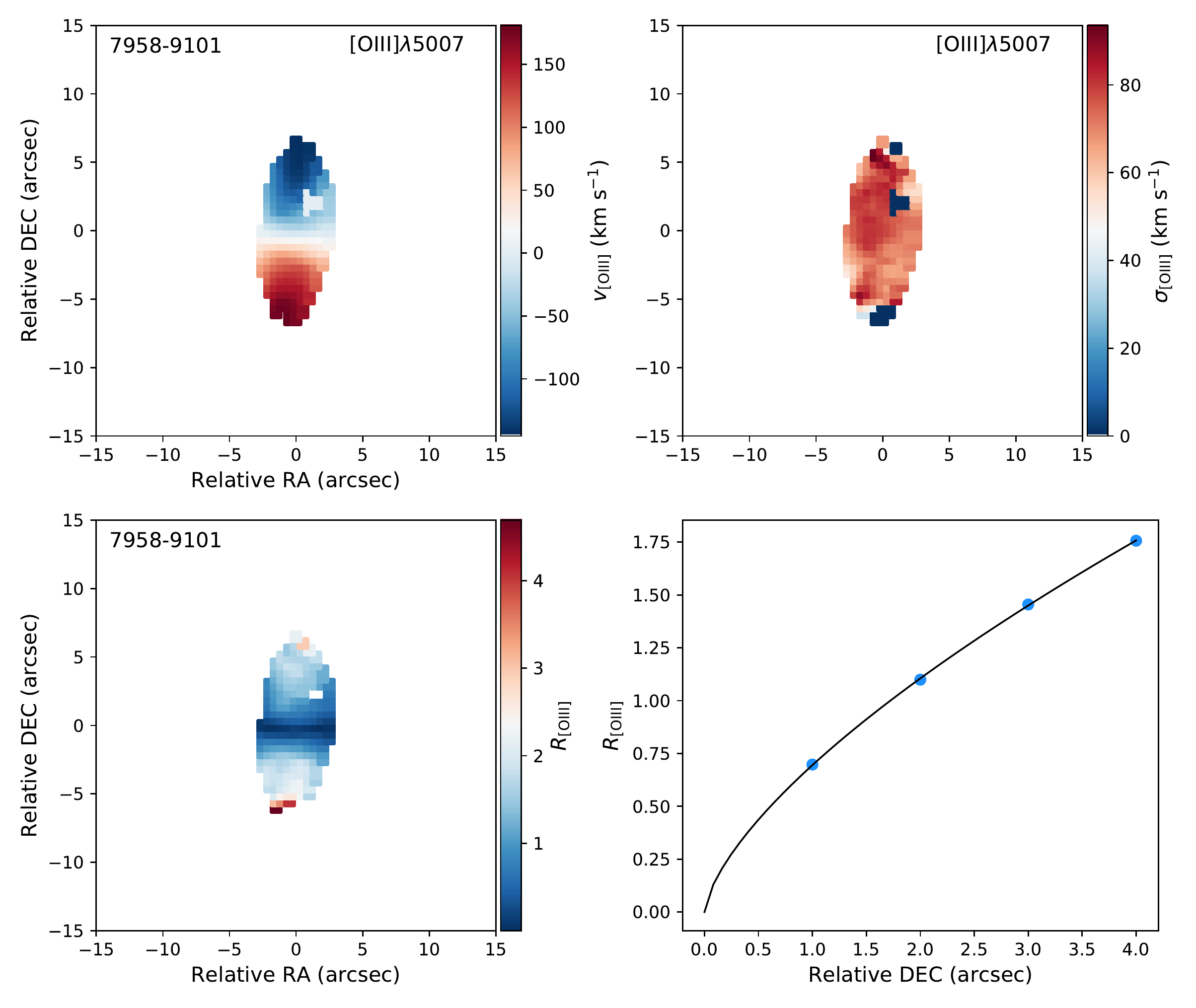}
\caption{MaNGA kinematics of the [OIII] gas for the AGN dwarf galaxy 7958-9101. From top to bottom, left to right: velocity, velocity dispersion, rotational support parameter ${\cal R}$, and power-law regression of the value of ${\cal R}$ as a function of the angular distance from the galaxy center.}
\label{fig:7958-9101}
\end{figure*}

We derive and report in Table \ref{table1} the parameter of rotational support at the central pixel (0.5'') for 14 dwarf galaxies with central AGN drawn from the AGN dwarf galaxy sample of \cite{Mezcua_2020_rot_sup}.
In Fig. \ref{fig:7958-9101} we show the kinematic maps of the [OIII] gas for the galaxy with MaNGA plateifu $7958-9101$, specifically its velocity, velocity dispersion, and rotational support parameter ${\cal R} = v_\mathrm{[OIII]}/\sigma_\mathrm{[OIII]}$. In the bottom right panel we show the results of a power-law regression of the value of ${\cal R}$ with the angular distance $\delta$ (in arcseconds) from the galactic centroid, obtaining the relation ${\cal R}(\delta) = (0.695 \pm 0.005) \, \delta^{(0.669 \pm 0.006)}$.

\begin{table*}[h!]
\centering
\caption{MaNGA plateifu, central velocity ($v_\mathrm{[OIII]}$) and velocity dispersion ($\sigma_\mathrm{[OIII]}$) of the [OIII] gas, central stellar velocity (v$_\mathrm{stellar}$) and stellar velocity dispersion ($\sigma_\mathrm{stellar}$), rotational support parameter of the [OIII] gas (${\cal R_\mathrm{gas}}$ = $v_\mathrm{[OIII]}$/$\sigma_\mathrm{[OIII]}$), rotational support parameter of the stars (${\cal R_\mathrm{stellar}}$ = v$_\mathrm{stellar}$/$\sigma_\mathrm{stellar}$).}
\label{table1}
\begin{tabular}{lcccccc}
\hline
\hline 
MaNGA & v$_{[OIII]}$  & $\sigma_{[OIII]}$ & v$_{stellar}$  & $\sigma_{stellar}$ & ${\cal R_\mathrm{gas}}$ & ${\cal R_\mathrm{stellar}}$ \\
plateifu & (km s$^{-1}$) &  (km s$^{-1}$) & (km s$^{-1}$) & (km s$^{-1}$) &  & \\
(1)       & (2)         & (3)             &     (4)         &       (5)       & (6)   & (7) \\
\hline
  7958-9101  &           11.00  &             79.54  &           14.71  &           117.88  &                     0.14  &                        0.12       \\
  8311-3701  &           10.23  &            103.93  &            5.37  &            98.84  &                     0.10  &                        0.05       \\
  8320-3704  &            2.76  &             59.27  &           15.97  &            54.93  &                     0.05  &                        0.29       \\
  8446-1901  &           44.76  &             99.22  &           42.74  &            65.55  &                     0.45  &                        0.65       \\
  8452-1901  &            0.52  &             48.98  &            5.25  &            65.18  &                     0.01  &                        0.08       \\
  8466-1901  &            3.61  &             77.77  &           17.21  &            82.29  &                     0.05  &                        0.21       \\
  8655-6103  &            1.55  &             33.21  &            1.97  &            45.28  &                     0.05  &                        0.04       \\
  8720-1901  &            0.89  &             71.72  &            1.83  &            60.92  &                     0.01  &                        0.03       \\
  8982-3703  &            9.65  &            149.31  &           15.63  &            47.31  &                     0.06  &                        0.33       \\
 8990-12705  &            5.94  &             82.85  &            3.34  &            68.40  &                     0.07  &                        0.05       \\
  8992-3702  &            5.63  &            106.21  &            7.17  &            70.04  &                     0.05  &                        0.10       \\
  9000-1901  &           22.69  &            120.84  &           81.74  &           174.26  &                     0.19  &                        0.47       \\
  9031-1902  &            4.80  &             68.32  &           12.81  &            89.33  &                     0.07  &                        0.14       \\
  9488-1901  &            9.83  &            122.41  &            7.23  &            79.93  &                     0.08  &                        0.09       \\
\hline
\hline
\end{tabular}
\end{table*}

\label{lastpage}
\end{document}